# Infrared thermotransmittance-based temperature field measurements in semitransparent media.

C. Bourges,[1] S. Chevalier,[2, a)] J. Maire,[1] A. Sommier,[1] C. Pradere,[3] and S. Dilhaire[4]

[1] *Univ. Bordeaux, CNRS, Bordeaux INP, I2M, UMR 5295, F-33400, Talence, France*

[2] *ENSAM, CNRS, Bordeaux INP, I2M, UMR 5295, F-33400 Talence, France*

[3] *EPSILON-ALCEN, F-33400 Talence, France*

[4] *Université de Bordeaux, CNRS, LOMA, UMR 5798, F-33400 Talence, France*

(Dated: 24 January 2023)

Temperature field contactless measurements in or at the surfaces of semitransparent media are a scientific challenge, as classical thermography techniques based on proper material emission cannot be used. In this work, an alternative method using infrared thermotransmittance for contactless imaging temperatures is proposed. To overcome the weakness of the measured signal, a lock-in acquisition chain is developed, and an imaging demodulation technique is used to retrieve the phase and amplitude of the thermotransmitted signal. These measurements, combined with an analytical model, enable the estimation of the thermal diffusivity and conductivity of an infrared semitransparent insulator (wafer of borofloat 33 glass) as well as the monochromatic thermotransmittance coefficient at 3.3 m. The obtained temperature fields are in good agreement with the model, and a detection limit of $\pm 2°$ C is estimated with this method. The results of this work open new opportunities in the development of advanced thermal metrology for semitransparent media.

[a)] Corresponding author stephane.chevalier@u-bordeaux.fr

## I. INTRODUCTION

Infrared (IR) semitransparent materials have been extensively studied in recent years because of their presence in many devices. For example, semiconductors, such as silicon or germanium, are used in batteries[1] and solar panels[2,3]. Some polymers[4] and glasses are also semitransparent and are mainly used in industry. To optimize the thermal performance of devices or monitor industrial processes[5], contactless measurement of temperature in materials is a major challenge. For thermal imaging measurements, the most widely used method is infrared thermography[6–9]. Although the method is perfectly adapted for opaque materials[10], it is not well-suited for semitransparent media. The latter absorb and transmit radiation from their surrounding environment, so discriminating the thermal signal of interest from parasitic radiation remains an obstacle.

Another approach relies on Raman spectroscopy[11,12], in which the peak position and width are temperature dependent. Raman thermometry is however limited to Raman active materials, which excludes many amorphous materials and metals. Furthermore, to avoid excessively long acquisition times, Raman thermometry is often performed with large temperature gradients. Finally, it is intrinsically a point by point technique. Other approaches are based on the thermal dependency of the fluorescence of specific particles. The medium must be inseminated with fluorescent nanoparticles or organic dyes[13]. However, the interactions between the particles and the material are not well known and possibly affect the thermal properties of the sample[14]. Thus, the development of a new contactless temperature measurement technique is of prime importance for resolving these issues.

An alternative method relies on the temperature dependency of the optical refractive index of materials[15–17]. Thermoreflectance[18–22] uses this characteristic to measure the thermal properties and surface temperatures of opaque and reflective materials. Analogous with thermoreflectance, thermotransmittance is the study of the thermally induced variation of transmitted light through a semitransparent medium. The variation in the transmitted light is related to the variation in temperature through the thermotransmittance coefficient $\kappa$. This parameter depends on the material, the thickness of the sample and the wavelength of the incident light, but only a few papers have reported the thermotransmittance coefficient values of some materials[23].

This coefficient can be determined either through optical property modeling or by exper-imental calibration. In the literature, no adequate theoretical model has been reported for estimating the thermotransmittance coefficient, and the infrared range has not been studied. The only other option is to measure the thermotransmittance coefficient of the sample under study. Works reporting such measurements are particularly scarce in the infrared domain and for semitransparent media. To address this challenge, a novel calibration method for the thermotransmittance coefficient is reported in this work.

Although thermotransmittance has been demonstrated to be well suited for temperature measurements in semitransparent materials, its sensitivity is weak in the infrared band[23], i.e., $\approx 10^{-4}$ K$^{-1}$. Therefore, careful attention should be given to electronic noise, thermal drift of devices, and parasitic environmental radiation in the signal acquisition chain. To resolve this major issue and improve the SNR, modulating the thermal excitation of the sample appears to be a promising approach. Appropriate demodulation enables the signal to be discarded at unwanted frequencies as a lock-in amplifier adapted for imaging systems[24,25]. To demonstrate the performance of the developed method for thermal properties and temperature measurements, an IR semitransparent wafer of borofloat 33 glass is used to create a temperature gradient within the material. Its thermal properties, namely, its thermal diffusivity and conductivity, are measured based on thermotransmittance and compared to values reported in the literature. Then, the thermotransmittance coefficient of the borofloat 33 wafer is calibrated to enable temperature measurement in this material.

## II. EXPERIMENTAL SETUP FOR THERMOTRANSMITTANCE MEASUREMENT

This section presents the experimental setup for measuring the thermotransmittance signal of a wafer of borofloat 33 glass. One of the great advantages of thermotransmittance is its proportionality to temperature variation. All of the presented study is based on this property.

### A. Experimental setup

The measurement process consists of detecting the transmitted intensity of an IR beam through the sample of interest with a camera. As the temperature of the sample varies, the

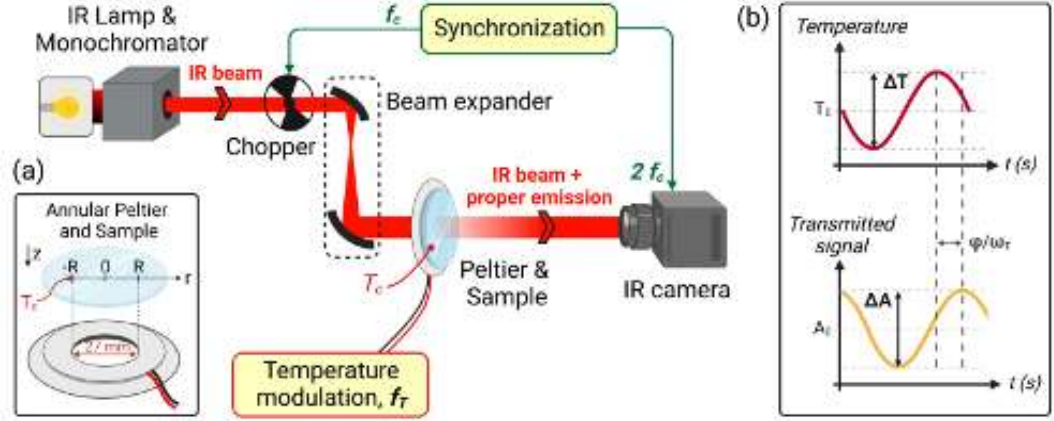

FIG. 1. (a) Illustration of the experimental setup: $f_T$ is the temperature modulation frequency, and $f_c$ is the chopper frequency. The inset in (a) presents a magnified view of the Peltier module and the sample (wafer of borofloat 33 glass). (b) Thermotransmittance principle: the temperature variation (red curve) induces a variation in the transmitted signal measured by the camera at the same frequency (yellow curve).

transmitted intensity also varies as follows:

$$\frac{\Delta A(t)}{A_0} = \kappa \Delta T(t) \qquad (1)$$

Equation 1 expresses the relationship among the averaged signal of the transmitted infrared beam $A_0$, the variation of the transmitted signal as a function of temperature $\Delta A(t)$, the temperature variation $\Delta T(t)$, and the time $t$.

The experimental setup for thermotransmittance measurement is presented in Fig. 1. The sample is heated by an annular Peltier module, with the temperature modulated around $T_0$ at the frequency $f_T$. A thermocouple combined with homemade LabView software enables temperature monitoring.

Upstream, the illumination wavelength is selected with a diffraction grating monochromator (Bentham Instruments, TMc300) from a stabilized infrared lamp (IR-Si217). After the monochromator, a beam expander adjusts the size of the infrared beam to the sample dimensions.

The infrared beam transmitted through the sample is measured by an infrared camera

(FLIR SC7000). The camera is composed of an indium-antimonide detector of $512 \times 640$ pixels with a pitch of 15 µm and a 50 mm focal length lens. The spectral wavelength range is $\lambda \in [2.5 - 5.5]$ µm, and the optical resolution in the experimental configuration is 107 µm/pixel.

The signal $S$ measured by the camera is a superposition[26] of the proper emission of scene $E$ and the infrared beam transmitted by sample $A$, i.e., $S = E + A$. The source chopping enables us to separate these contributions: the opening of the chopper provides $S$, and its closing provides $E$. The camera acquisition frame rate is synchronized at twice the chopper frequency $2f_c$. Finally, $A$ is retrieved by subtracting two successive frames (see Fig.2 (b)). To prevent a change in the proper emission between the subtracted images, the chopper frequency, $f_c = 25$ Hz, is set much larger than the thermal frequency, $f_T = 5$ mHz.

To demonstrate this method, we use a semitransparent insulator in the spectral range of the camera. The medium selected for study is a wafer of borofloat 33 glass[27]. The thickness of the sample is measured with a micrometric caliper, $e = 515 \pm 10$ µm, and its diameter is $d = 50.8$ mm.

### B. Signal processing

To retrieve the amplitude $\Delta A$ and phase $\varphi$ of the modulated thermotransmitted signal $A(t)$, the four-image method is implemented[28], as described in Equations 2 and 3.

The signal-to-noise ratio is improved by removing the signal at unwanted frequencies $f \neq f_T$. To further reduce the noise, two strategies are implemented. First, measurements are repeated on several periods and averaged. Second, the camera is triggered to record four sets of frames per period (Fig.2 (b)).

Each image $I_{mi}$ used in Eqs. 2 and 3 is an average of the corresponding set of frames after proper emission subtraction.

$$\Delta A = \sqrt{(I_{m1} - I_{m3})^2 + (I_{m2} - I_{m4})^2} \quad (2)$$

$$\varphi = \tan^{-1}\left(\frac{I_{m4} - I_{m2}}{I_{m1} - I_{m3}}\right) + \varphi_0 \quad (3)$$

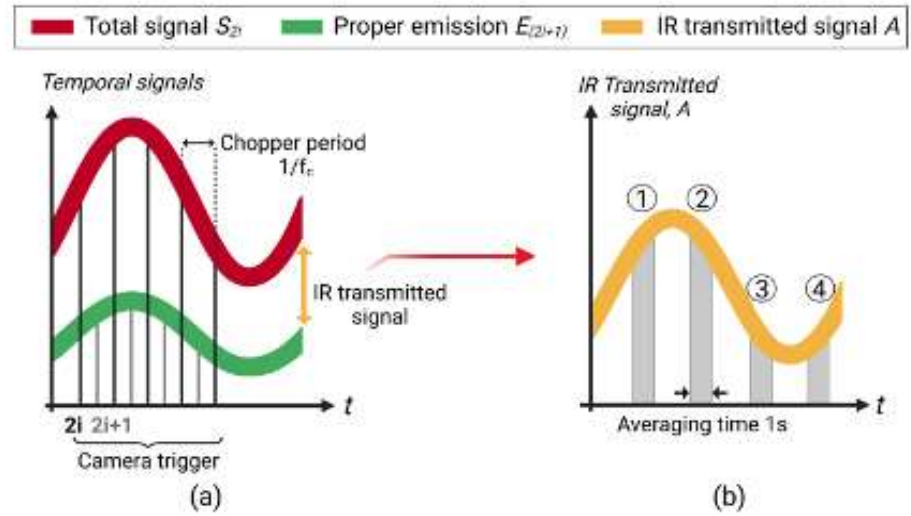

FIG. 2. (a) Temporal signals of one pixel, measured by the camera at a frequency of $2f_c$. The proper emission E is captured at odd frames $2i+1$, and the total signal $S = E + A$ is captured at even frames $2i$. Subtracting the proper emission from the total signal yields the transmitted IR signal $A$. (b) Illustration of the four sets of frames used for amplitude and phase measurements of a sine wave temperature modulation.

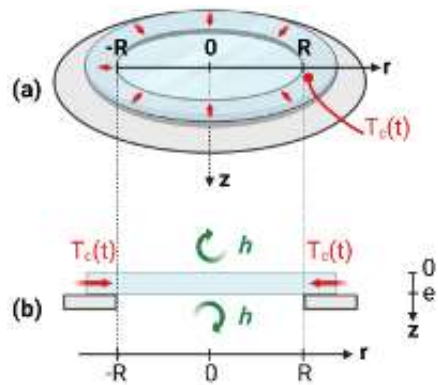

FIG. 3. (a) Three-dimensional view of the Peltier module and the sample heated to a modulated temperature $T_c(t)$. (b) Cross-sectional view of the borofloat 33 glass wafer, with the associated boundary conditions.

### C. Heat transport modeling

The heat transfer across the system is modeled to validate the experimental thermo-transmittance measurements and to estimate its thermal properties. We consider a sample of radius $R$ and thickness $e$, with a thermal diffusivity $a$ (m$^2$/s) and a thermal conductivity $k$ (W/m/K). At the edges (Fig. 3), namely, where $r = R$ and $r = -R$, the temperature is set to the modulated temperature $T_c(t)$. For symmetry reasons, at the position $r = 0$, the flux is assumed to be equal to zero, and convection losses are imposed at $z = 0$ and $z = e$, with $h$ being the convection coefficient. The heat equation in cylindrical coordinates and the associated boundary conditions are expressed in Eqs. 4 to 9.

$$\begin{cases} \dfrac{\partial^2 T(r,z,t)}{\partial r^2} + \dfrac{1}{r}\dfrac{\partial T(r,z,t)}{\partial r} + \dfrac{\partial^2 T(r,z,t)}{\partial z^2} = \dfrac{1}{a}\dfrac{\partial T(r,z,t)}{\partial t} & (4) \\[6pt] -k\left.\dfrac{\partial T(r,z,t)}{\partial z}\right|_{z=0} = -h(T(r,z=0,t) - T_0) & (5) \\[6pt] -k\left.\dfrac{\partial T(r,z,t)}{\partial z}\right|_{z=e} = h(T(r,z=e,t) - T_0) & (6) \\[6pt] -k\left.\dfrac{\partial T(r,z,t)}{\partial r}\right|_{r=0} = 0 & (7) \\[6pt] T(r=R,z,t) = T(r=-R,z,t) = T_c(t) = T_0 + \dfrac{\Delta T}{2}\sin(\omega_T t) & (8) \\[6pt] T(r,z,t=0) = T_0 & (9) \end{cases}$$

The Biot number of the wafer of borofloat 33 is $B_i = he/k \approx 5.10^{-3} \ll 1$, with $k \approx 1\,\text{W/m/K}$ and $h \approx 10\,\text{W/m}^2/\text{K}$ for the vertical configuration[29] used in our setup. Therefore, the temperature along the z-dimension should be uniform, and the partial derivative with respect to z can be linearized.

$$\begin{aligned} \dfrac{\partial^2 T(r,z,t)}{\partial z^2} &\approx \dfrac{\left.\dfrac{\partial T(r,z,t)}{\partial z}\right|_{z=e} - \left.\dfrac{\partial T(r,z,t)}{\partial z}\right|_{z=0}}{e} \\ &\approx -\dfrac{2h}{ke}(T(r,z=e,t) - T_0) \end{aligned} \quad (10)$$

To solve the system, the temperature is decomposed into a constant and a term depending on the pulsation $\omega_T = 2\pi f_T$:

$$T(r,t) = T_0 + \theta(r,\omega_T)e^{i\omega_T t} \quad (11)$$

The previous system of equations (from 4 to 9) becomes

$$\begin{cases} \dfrac{d^2\theta(r,\omega_T)}{dr^2} + \dfrac{1}{r}\dfrac{d\theta(r,\omega_T)}{dr} - \dfrac{2h}{ke}\theta(r,\omega_T) = \dfrac{i\omega_T}{a}\theta(r,\omega_T) & (12) \\ -k\left.\dfrac{d\theta(r,\omega_T)}{dr}\right|_{r=0} = 0 & (13) \\ \theta(r=R,\omega_T) = \theta_c(\omega_T) & (14) \end{cases}$$

The solution of the system is a modified Bessel function of the first kind $I_0$[30]. Finally, the complex temperature field in the sample as a function of the position $r$ and the frequency $f_T = 2\pi/\omega_T$ is expressed as follows:

$$\theta(r,\omega_T) = \theta_c(\omega_T)\dfrac{I_0\left(r\sqrt{H + i\frac{\omega_T}{a}}\right)}{I_0\left(R\sqrt{H + i\frac{\omega_T}{a}}\right)} \quad (15)$$

where $H = 2h/ke$ is the loss factor. The amplitude and phase of the complex temperature $\theta(r,\omega_T)$ are shown in Fig.4 (a) and (b).

### D. Sensitivity analysis

A sensitivity study (Fig.4 (c) and (d)) is performed to determine the influence of thermal properties on the measured amplitude and phase, and to adapt the minimization algorithm (see equation 19). A variation of 10% is applied to the thermal diffusivity $a$ and the term $H$. The sensitivities of the amplitude $S_\theta$ and phase $S_\varphi$ to $a$ and $H$ are expressed in detail by Eqs.16 and 17.

$$S_{\theta a} = a\dfrac{\partial|\theta(a,H)|}{\partial a} \qquad S_{\theta H} = H\dfrac{\partial|\theta(a,H)|}{\partial H} \quad (16)$$

$$S_{\varphi a} = a\dfrac{\partial\varphi(a,H)}{\partial a} \qquad S_{\varphi H} = H\dfrac{\partial\varphi(a,H)}{\partial H} \quad (17)$$

According to the characteristics of the material and the geometry of the setup, the values used in the model are $a = 7 \times 10^{-7}\,\mathrm{m^2/s}$, $H = 3.45 \times 10^4\,\mathrm{m^{-2}}$, $k = 1\,\mathrm{W/m/K}$, $R = 13.5\,\mathrm{mm}$, $e = 0.5\,\mathrm{mm}$, $\theta_c = 20\,^\circ\mathrm{C}$, and $f_T = 5\,\mathrm{mHz}$. Figure 4 shows that the amplitude is more sensitive to convective losses, in contrast to the phase, which is more sensitive to the thermal diffusivity. The results of the sensitivity study demonstrate the importance of using both the amplitude and phase to properly estimate the thermal properties of the material.

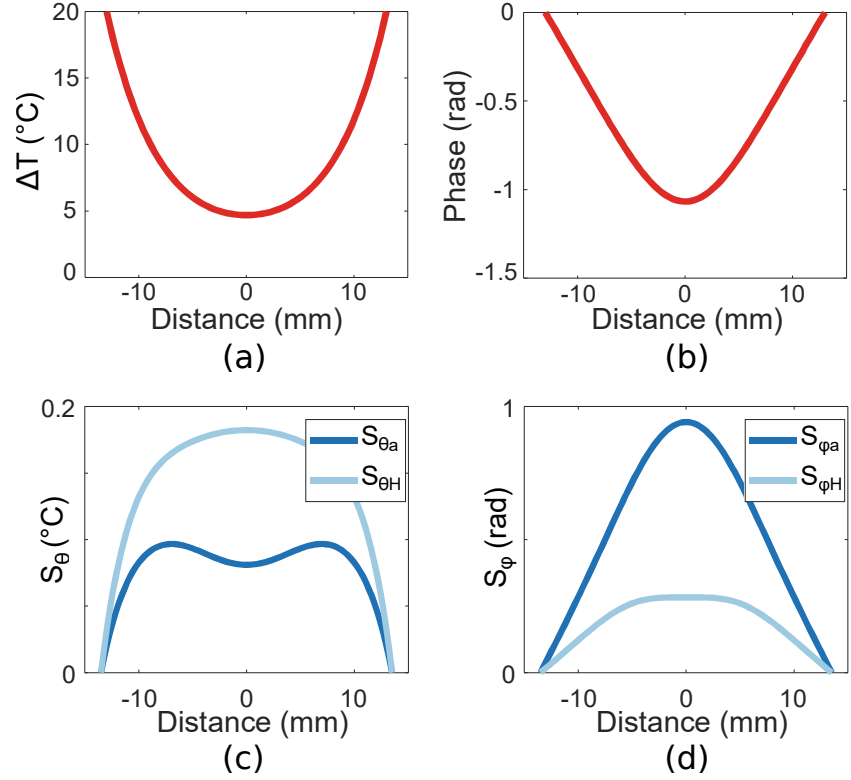

FIG. 4. Amplitude (a) and phase (b) of the expected temperature variation as functions of the distance $r$. Amplitude (c) and phase (d) sensitivities as functions of $a$ and $H$.

## III. RESULTS AND DISCUSSION

At any given wavelength, the measured signal is directly proportional to the amount of IR light detected by the camera. Thus, the chosen wavelength corresponds to the maximum transmission of the borofloat sample, which is $\lambda = 3300$ nm. The sample is heated at its edges at the modulated temperature $T_c(t) = T_0 + (\Delta T/2)\sin(\omega_T t)$, with $T_0 = 30\,°C$, $\Delta T = 20\,°C$ and $\omega_T = 2\pi f_T$. To ensure that the heat diffuses to the center of the wafer, the characteristic thermal frequency is roughly estimated to be $f \approx a/d^2 \approx 7 \times 10^{-7}/(0.013)^2 \approx 5$ mHz, and the modulation frequency is then set to $f_T = 5$ mHz.

The measured amplitude and phase fields of the thermotransmitted signal are presented in Fig. 5. As expected, the fields are axisymmetric, and the amplitude is higher at the edges, where the sample is heated. The phase is equal to zero at the edges, in phase with the setpoint temperature. To compare the measurements to the model, each pixel position $(x, y)$ is converted to a distance $r = \sqrt{(x-x_0)^2 + (y-y_0)^2}$ from the center $(x_0, y_0)$ of the sample. All corresponding data points are presented in Fig.5 (c) for the amplitude and in Fig.5 (d) for the phase. The noticeable outliers originate from scratches on the sample and unevenness on the edges of the Peltier module, but the results are not affected.

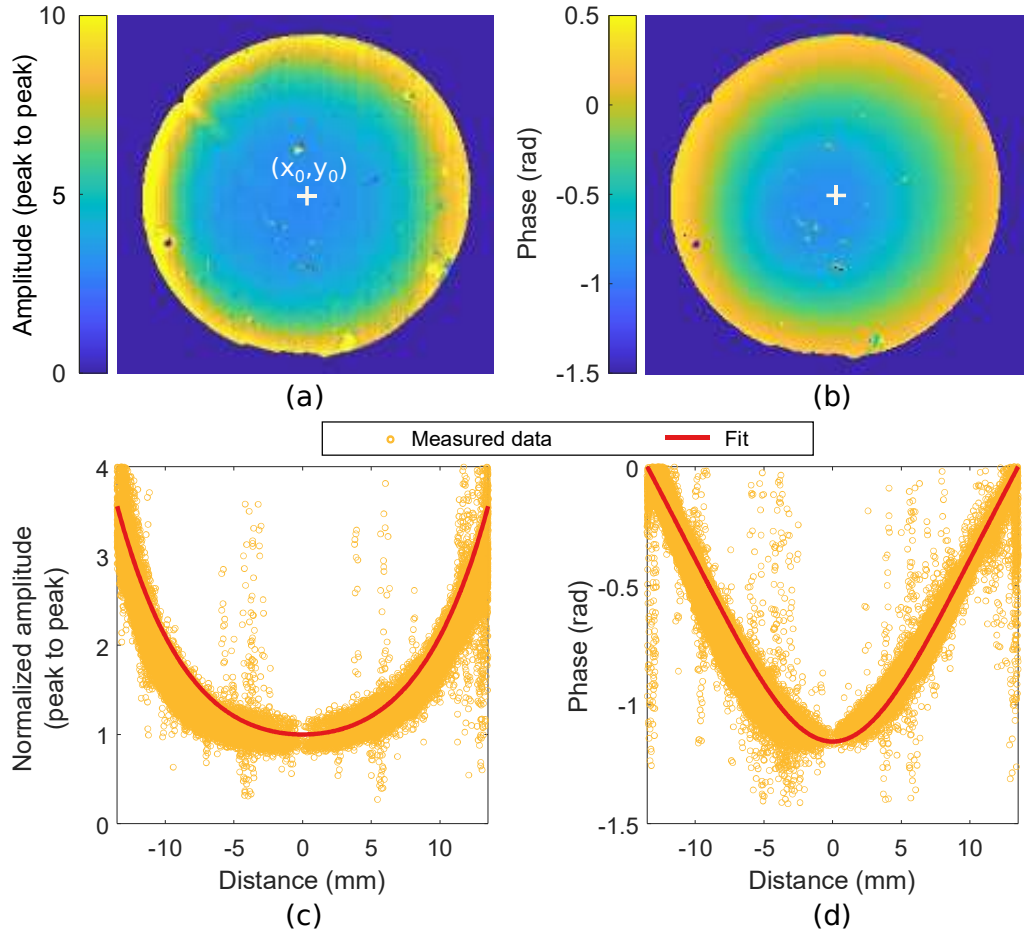

FIG. 5. (a) Amplitude field measurement of the thermotransmitted signal at $f_T = 5\,\text{mHz}$ and $\lambda = 3.3\,\mu\text{m}$. The image center $(x_0, y_0)$ is marked by the white cross. (b) Phase field measurement of the thermotransmitted signal. (c) Amplitude normalized at $r = 0$ and (d) phase as a function of the distance to the center $r$. As the thermotransmittance signal is in phase opposition with the temperature, $\pi$ was subtracted from the data. Yellow circles: data corresponding to $5 \times 10^4$ pixels. Red curve: minimization result (see Eq.19 and Tab.I).

To validate the thermotransmittance method for temperature measurement, the thermal properties of the material are estimated by minimizing the norm between the model and the measurements. First, they are normalized by the data at the position $r = 0$ (see Eq. 18) to remove from the equation the thermotransmittance coefficient and the temperature variation at the edge $\Delta T$:

$$\frac{\frac{\Delta A}{A_0}(r, \omega_T)}{\frac{\Delta A}{A_0}(0, \omega_T)} = \frac{|\theta(r, \omega_T)|}{|\theta(0, \omega_T)|} \qquad (18)$$

This step reduces the number of free parameters and increases the estimation accuracy of the thermal properties of the material. To estimate the free parameters a and H, a Matlab derivative-free algorithm (simplex algorithm from the $fminsearch$ subroutine) was used to minimize the error between the model and the measurements. The cost function which measures this error was chosen based on the amplitude and phase of the signal as defined in the following equation:

$$J = \left\| \frac{|\theta(r, \omega_T)|}{|\theta(0, \omega_T)|} - \frac{\Delta A/A_0(r, \omega_T)}{\Delta A/A_0(0, \omega_T)} \right\|^2 + \|\varphi_{model}(r, \omega_T) - \varphi_{measured}(r, \omega_T)\|^2 \qquad (19)$$

Considering known values[27] of the mass density, $\rho = 2200 \, \text{kg/m}^3$, and the specific heat capacity, $c_p = 800 \, \text{J/kg/K}$, of the material, the thermal conductivity, $k = a\rho c_p$, and the convection coefficient depending on the experimental conditions, $h = Hke/2$, are derived and presented in Table I. The relative error is 8% for the thermal diffusivity and conductivity measurement and is close to 20% for the convection coefficient, which is satisfactory for thermal property measurement. The expected thermal properties[27] are $k \approx 1.2 \, \text{W/m/K}$ and $a \approx 7 \times 10^{-7} \, \text{m}^2/\text{s}$, which are consistent with the obtained measurements. These results validate the use of thermotransmittance for thermal property estimation in semitransparent materials.

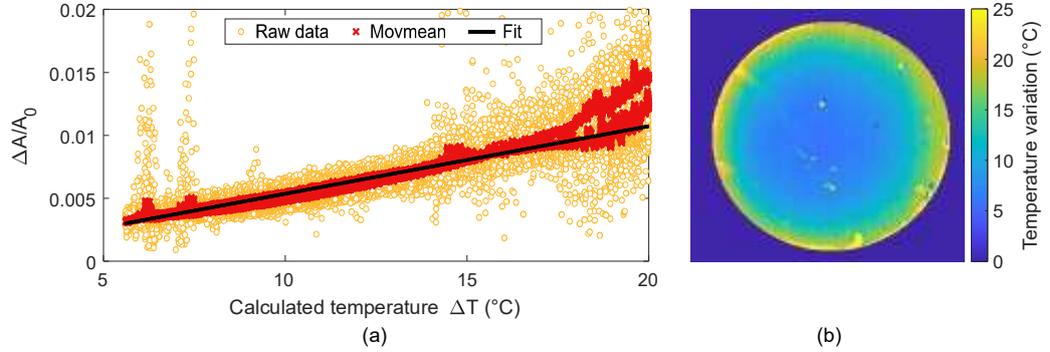

FIG. 6. (a) Thermotransmitted signal as a function of the calculated temperature variation. Yellow circles: data corresponding to $5 \times 10^4$ pixels. Red crosses: moving averages over 50 points. Solid line: linear regression. (b) Measured map of the temperature variation, $\Delta T$.

TABLE I. Estimation of thermal properties of the borofloat wafer

| Parameter | Estimated value |
|---|---|
| Thermal diffusivity $a$ | $a = (7.4 \pm 0.6) \times 10^{-7}$ m$^2$/s |
| Loss factor, $H$ | $H = (2.9 \pm 0.5) \times 10^4$ m$^{-2}$ |
| Thermal conductivity, $k$ | $k = 1.3 \pm 0.1$ W/m/K |
| Convection coefficient, $h$ | $h = 9.5 \pm 1.8$ W/m$^2$/K |

Finally, the thermotransmittance coefficient is estimated from the ratio between the temperature calculated from the model and the measured amplitude $\Delta A/A_0$, as presented in Fig. 6 (a). Therefore, the slope of the linear fit provides the thermotransmittance coefficient $\kappa_{borofloat} = -(5.2 \pm 0.2) \times 10^{-4}$K$^{-1}$ at $\lambda = 3300$ nm for the borofloat wafer of thickness 515 µm. The minus sign comes from the phase opposition between the temperature and the thermotransmitted signal: the higher the temperature is, the more opaque the sample.

With knowledge of the thermotransmittance coefficient, temperature variation measurements can be performed in semitransparent media (see Fig. 6 (b)). In the current implementation of the setup (with camera electronic noise and the selected number of averaged periods), the sensitivity is $\sigma_{\Delta A/A_0} = 0.0012$. The noise converted into temperature with $\kappa$ is $\sigma_T \approx 2\,°$C. According to Fig. 6 (b), the minimal measured temperature variation is $\Delta T_{min} = 5\,°C > \sigma_T$. Therefore, the presented results are not impacted by the thermal resolution under the chosen experimental conditions.

The second interesting experimental finding is the minimal acquisition time to heat the sample $\tau \approx d^2/a$. For the chosen glass wafer, $\tau \approx 200\,s$, which corresponds to $f_T = 5\,mHz$. Reducing $\tau$ means working with a smaller sample: for a micrometric sample with the same thermal diffusivity, $\tau \approx (100^{-6})^2/a \approx 10\,ms$. To optimize the thermotransmittance measurement performance for such a sample, the heating system must be adjusted according to the dimensions of the material (using laser heating, the Joule effect, etc.). Therefore, the thermotransmittance method should provide temperature variation measurements at any size scale, requiring an adjustment of the heating conditions. This paper presents an example of a millimeter system, and future work will enable the development of the method for a smaller system.

To measure the temperature in a new sample, the thermotransmittance coefficient of the material must be estimated first. As discussed in the introduction, there is no adequate model for estimating the thermotransmittance coefficient. Drude theory[31,32] provides a working hypothesis for metals from which one can develop a more general model. The availability of an experimental database of the thermotransmittance coefficient values of diverse materials with different thicknesses would facilitate the selection of the appropriate hypothesis, depending on the nature of the material (metal[33], semiconductor[34], or dielectric).

The current configuration provides the average temperature of the sample over its thickness. To address in-situ temperature measurement in thick samples, three-dimensional (3D) thermotransmittance measurement has to be performed. Several tomography methods will be investigated such as confocal imaging[35] or Radon tomography[36].

**CONCLUSIONS**

The study demonstrates that imaging-modulated thermotransmittance is a powerful method for thermal properties and temperature measurement in semitransparent materials: the signal is directly proportional to the variation of the temperature of the material.

The thermotransmittance coefficient $\kappa$ of borofloat 33 was measured, as well as the thermal properties. Moreover, thermal modulation strongly increases the SNR of the measurement, which was formerly limited by electronic noise, parasitic signals or thermal drift of the devices.

In the current setup, the time of the experiment is approximately 200 s due to the low thermal frequency. To study transient phenomena, an adaptation of the current setup is necessary to lower the acquisition time. Moreover, reducing the size of the system and modifying the heat source (e.g., using a laser or the Joule effect) should enable the study of micro devices such as microsupercapacitors.

Finally, thermotransmittance is expected to be a powerful method for quantitative thermal tomography in semitransparent media. Future works will focus on improving the current setup for 3D-field temperature measurements.

## DATA AVAILABILITY STATEMENT

The data that support the findings of this study are available from the corresponding author upon reasonable request.

## REFERENCES

[1] K. Peng, J. Jie, W. Zhang, and S. T. Lee, "Silicon nanowires for rechargeable lithium-ion battery anodes," Applied Physics Letters **93**, 1–4 (2008).

[2] K. Niazi, H. A. Khan, and F. Amir, "Hot-spot reduction and shade loss minimization in crystalline-silicon solar panels," Journal of Renewable and Sustainable Energy **10** (2018), 10.1063/1.5020203.

[3] J. Shin, J. Park, and N. Park, "A method to recycle silicon wafer from end-of-life photovoltaic module and solar panels by using recycled silicon wafers," Solar Energy Materials and Solar Cells **162**, 1–6 (2017).

[4] J. Gieseler, A. Adibekyan, C. Monte, and J. Hollandt, "Apparent emissivity measurement of semi-transparent materials part 1: Experimental realization," Journal of Quantitative Spectroscopy and Radiative Transfer **257**, 107316 (2020).

[5] M. Speka, S. Matteï, M. Pilloz, and M. Ilie, "The infrared thermography control of the laser welding of amorphous polymers," NDT and E International **41**, 178–183 (2008).

[6] X. Maldague and S. Marinetti, "Pulse phase infrared thermography," Journal of Applied Physics **79**, 2694–2698 (1996).

[7] F. Cernuschi, A. Russo, L. Lorenzoni, and A. Figari, "In-plane thermal diffusivity evaluation by infrared thermography," Review of Scientific Instruments **72**, 3988–3995 (2001).

[8] G. M. Carlomagno and G. Cardone, *Experiments in Fluids*, Vol. 49 (2010) pp. 1187–1218.

[9] R. Usamentiaga, P. Venegas, J. Guerediaga, L. Vega, J. Molleda, and F. G. Bulnes, "Infrared thermography for temperature measurement and non-destructive testing," Sensors (Switzerland) **14**, 12305–12348 (2014).

[10] C. Meola, S. Boccardi, and G. M. Carlomagno, "Measurements of very small temperature variations with LWIR QWIP infrared camera," Infrared Physics and Technology **72**, 195–203 (2015).

[11] S. P. Nikitin, C. Manka, J. Grun, and J. Bowles, "A technique for contactless measurement of water temperature using Stokes and anti-Stokes comparative Raman spectroscopy," Review of Scientific Instruments **83** (2012), 10.1063/1.3685613.

[12] D. S. Moore and S. D. McGrane, "Raman temperature measurement," Journal of Physics: Conference Series **500** (2014), 10.1088/1742-6596/500/19/192011.

[13] P. Löw, B. Kim, N. Takama, and C. Bergaud, "High-spatial-resolution surface-temperature mapping using fluorescent thermometry," Small **4**, 908–914 (2008).

[14] F. Wang, Y. Han, and N. Gu, "Cell Temperature Measurement for Biometabolism Monitoring," ACS Sensors **6**, 290–302 (2021).

[15] W. J. Tropf and M. E. Thomas, "Infrared refractive index and thermo-optic coefficient measurement at APL," Johns Hopkins APL Technical Digest (Applied Physics Laboratory) **19**, 293–297 (1998).

[16] W. Wang, Y. Yu, Y. Geng, and X. Li, "Measurements of thermo-optic coefficient of standard single mode fiber in large temperature range," 2015 International Conference on Optical Instruments and Technology: Optical Sensors and Applications

[17] H. H. Li, "Refractive index of silicon and germanium and its wavelength and temperature derivatives," Journal of Physical and Chemical Reference Data **9**, 561–658 (1980).

[18] S. Dilhaire, S. Grauby, and W. Claeys, "Calibration procedure for temperature measurements by thermoreflectance under high magnification conditions,"

Applied Physics Letters **84**, 822–824 (2004).

[19]D. U. Kim, K. S. Park, C. B. Jeong, G. H. Kim, and K. S. Chang, "Quantitative temperature measurement of multi-layered semiconductor devices using spectroscopic thermoreflectance microscopy," Optics Express **24**, 13906 (2016).

[20]D. G. Cahill, "Analysis of heat flow in layered structures for time-domain thermoreflectance," Review of Scientific Instruments **75**, 5119–5122 (2004).

[21]M. Farzaneh, K. Maize, D. Lüerßen, J. A. Summers, P. M. Mayer, P. E. Raad, K. P. Pipe, A. Shakouri, R. J. Ram, and J. A. Hudgings, "Ccd-based thermoreflectance microscopy: principles and applications," Journal of Physics D: Applied Physics **42**, 143001 (2009).

[22]J. Christofferson and A. Shakouri, "Thermoreflectance based thermal microscope," Review of Scientific Instruments **76**, 024903–1–024903–6 (2005).

[23]C. Pradere, M. Ryu, A. Sommier, M. Romano, A. Kusiak, J. L. Battaglia, J. C. Batsale, and J. Morikawa, "Non-contact temperature field measurement of solids by infrared multispectral thermotransmittance," Journal of Applied Physics **121** (2017), 10.1063/1.4976209.

[24]R. Mulaveesala and S. Tuli, "Theory of frequency modulated thermal wave imaging for nondestructive subsurface defect detection," Applied Physics Letters **89**, 1–4 (2006).

[25]O. Breitenstein, M. Langenkamp, F. Altmann, D. Katzer, A. Lindner, and H. Eggers, "Microscopic lock-in thermography investigation of leakage sites in integrated circuits," Review of Scientific Instruments **71**, 4155–4160 (2000), https://aip.scitation.org/doi/pdf/10.1063/1.1310345.

[26]T. Lafargue-Tallet, R. Vaucelle, C. Caliot, A. Aouali, E. Abisset-Chavanne, A. Sommier, R. Peiffer, and C. Pradere, "Active thermo-reflectometry for absolute temperature measurement by infrared thermography on specular materials," Scientific Reports **12**, 1–19 (2022).

[27]SCHOTT Technical Glass Solutions GmbH, "Schott Borofloat 33," , 1–32 (2009).

[28]G. Busse, D. Wu, and W. Karpen, "Thermal wave imaging with phase sensitive modulated thermography," Journal of Applied Physics **71**, 3962–3965 (1992).

[29]S. W. Churchill and H. H. Chu, "Correlating equations for laminar and turbulent free convection from a vertical plate," International Journal of Heat and Mass Transfer **18**, 1323–1329 (1975).

[30]D. W. Hahn and M. N. Ozisik, *Heat Conduction, third edition* (Wiley and Sons, 2012).

[31] M. A. Ordal, R. J. Bell, R. W. Alexander, L. L. Long, and M. R. Querry, "Optical properties of fourteen metals in the infrared and far infrared: Al, Co, Cu, Au, Fe, Pb, Mo, Ni, Pd, Pt, Ag, Ti, V, and W," Applied Optics **24**, 4493 (1985).

[32] A. Block, M. Liebel, R. Yu, M. Spector, Y. Sivan, F. J. García De Abajo, and N. F. Van Hulst, "Tracking ultrafast hot-electron diffusion in space and time by ultrafast thermomodulation microscopy," Science Advances **5**, 1–8 (2019), 1809.10591.

[33] R. Rosei and D. W. Lynch, "Thermomodulation spectra of Al, Au, and Cu," Physical Review B **5**, 3883–3894 (1972).

[34] G. Ghosh, "Temperature dispersion of refractive indices in semiconductors," Journal of Applied Physics **79**, 9388–9389 (1996).

[35] J. Jonkman, "Tutorial: guidance for quantitative confocal microscopy," Nature Protocols **15**, 1585–1611 (2020).

[36] A. G. Ramm and A. I. Katsevich, *Radon transform and local tomography* (CRC press, 2020).

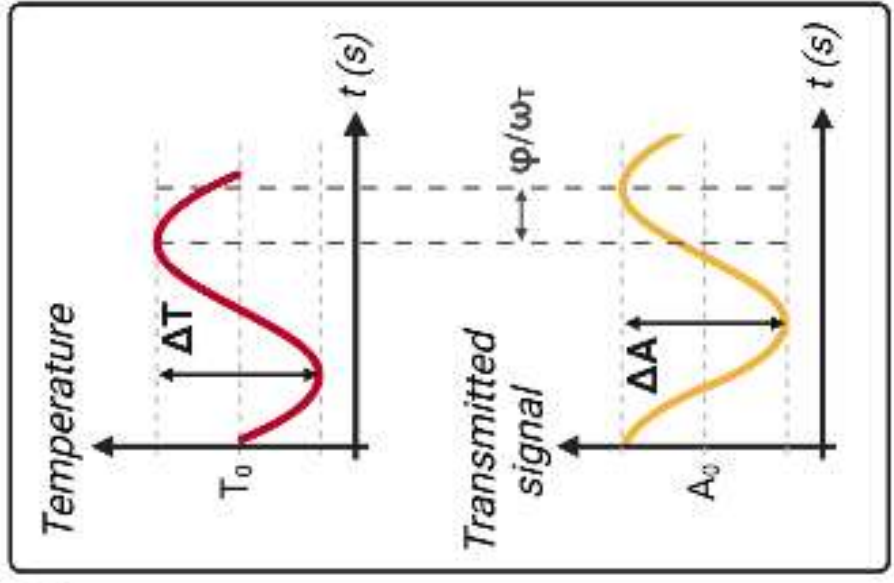
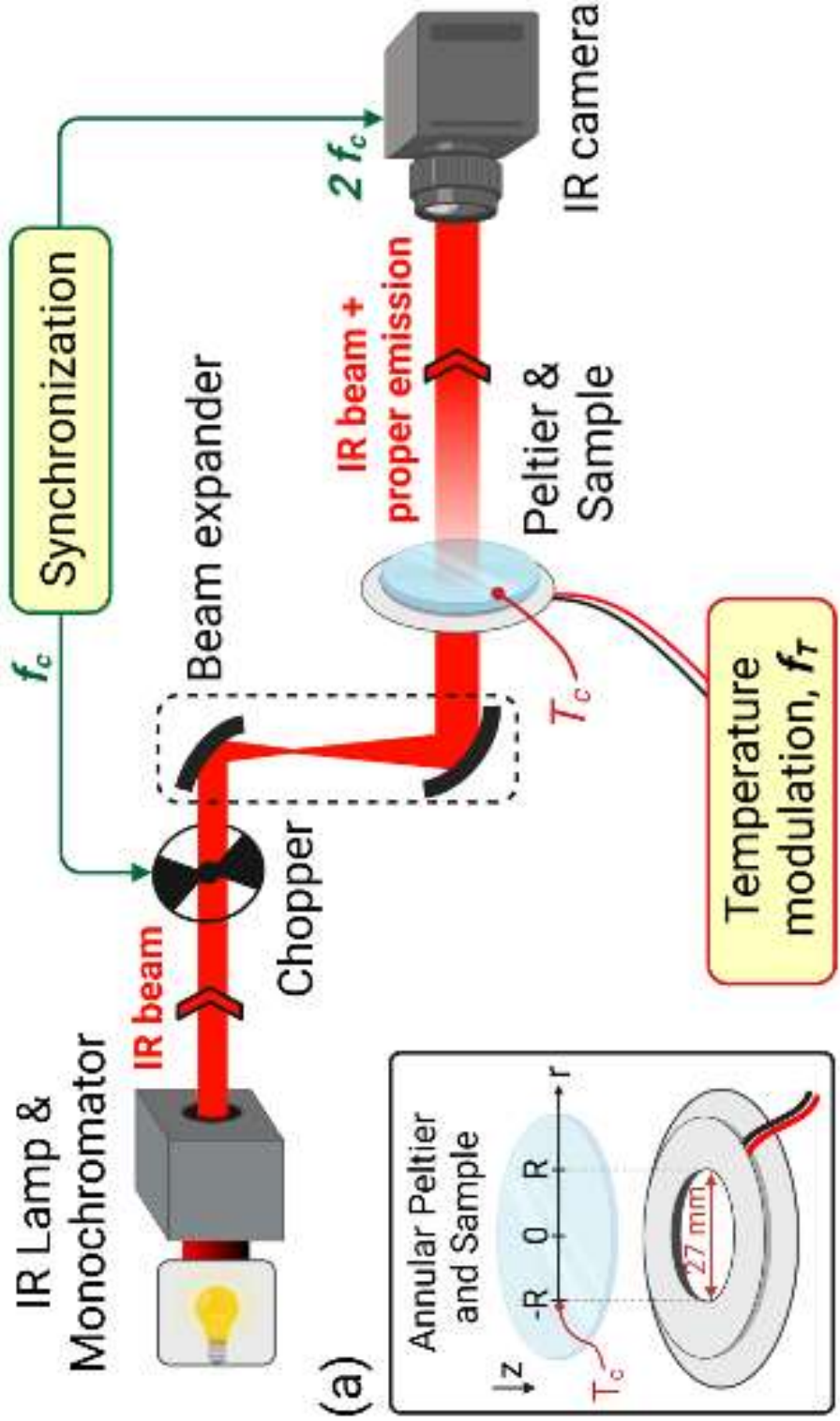

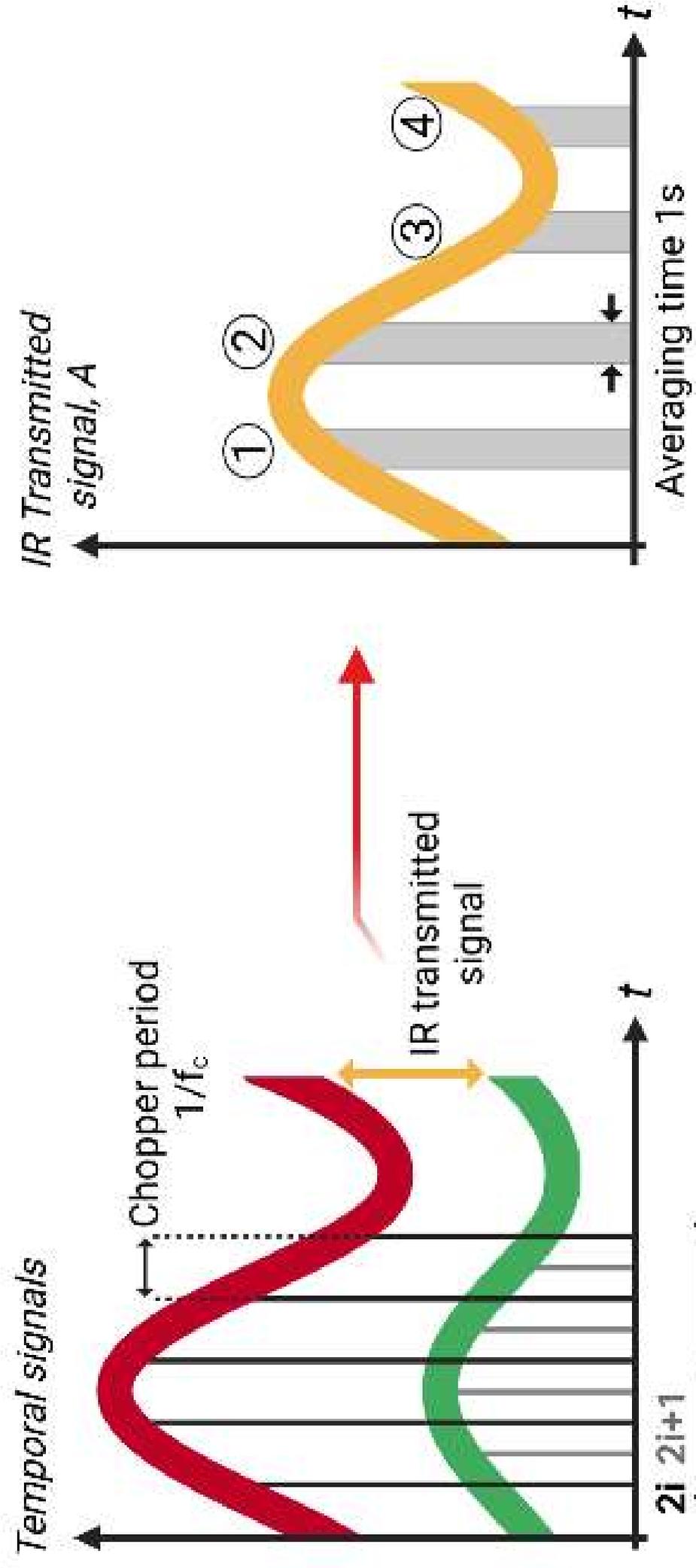

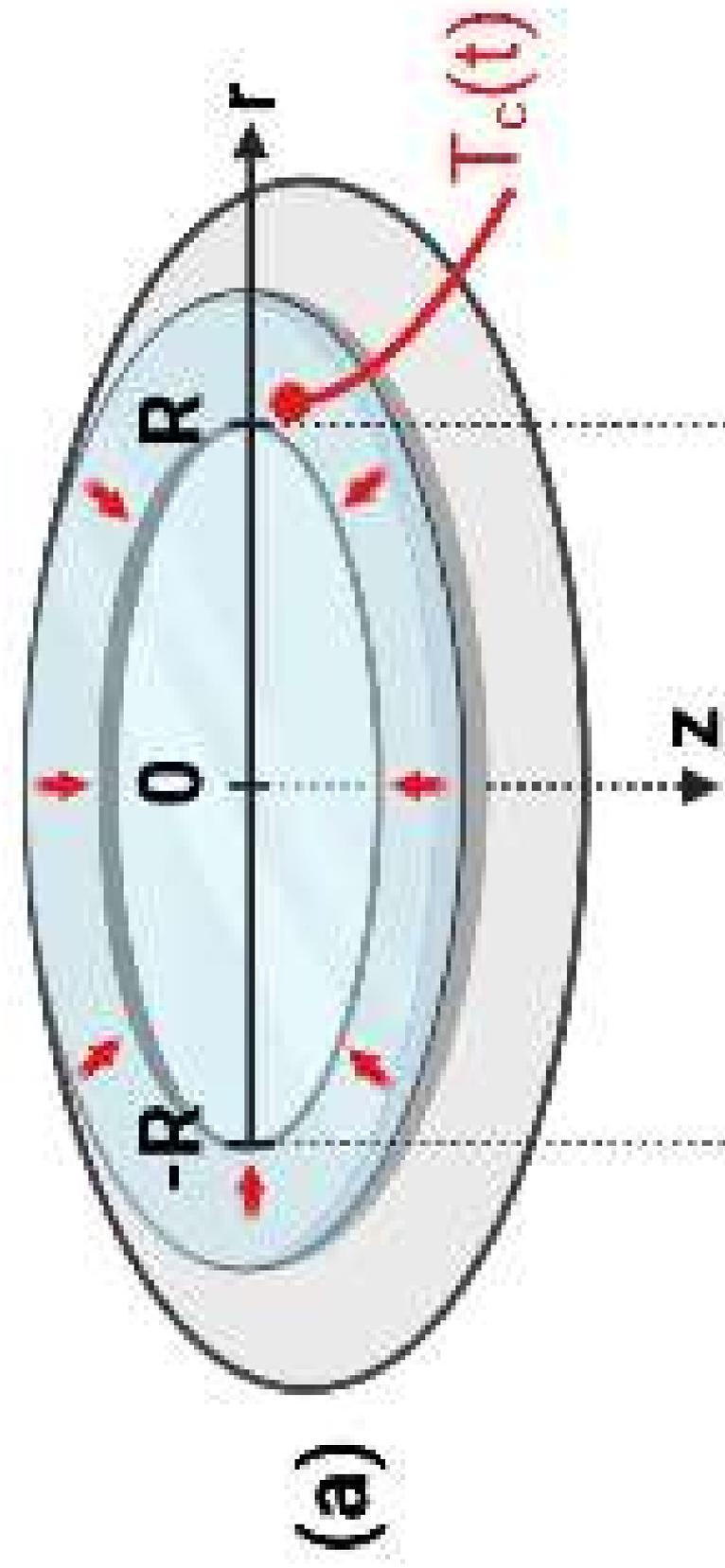
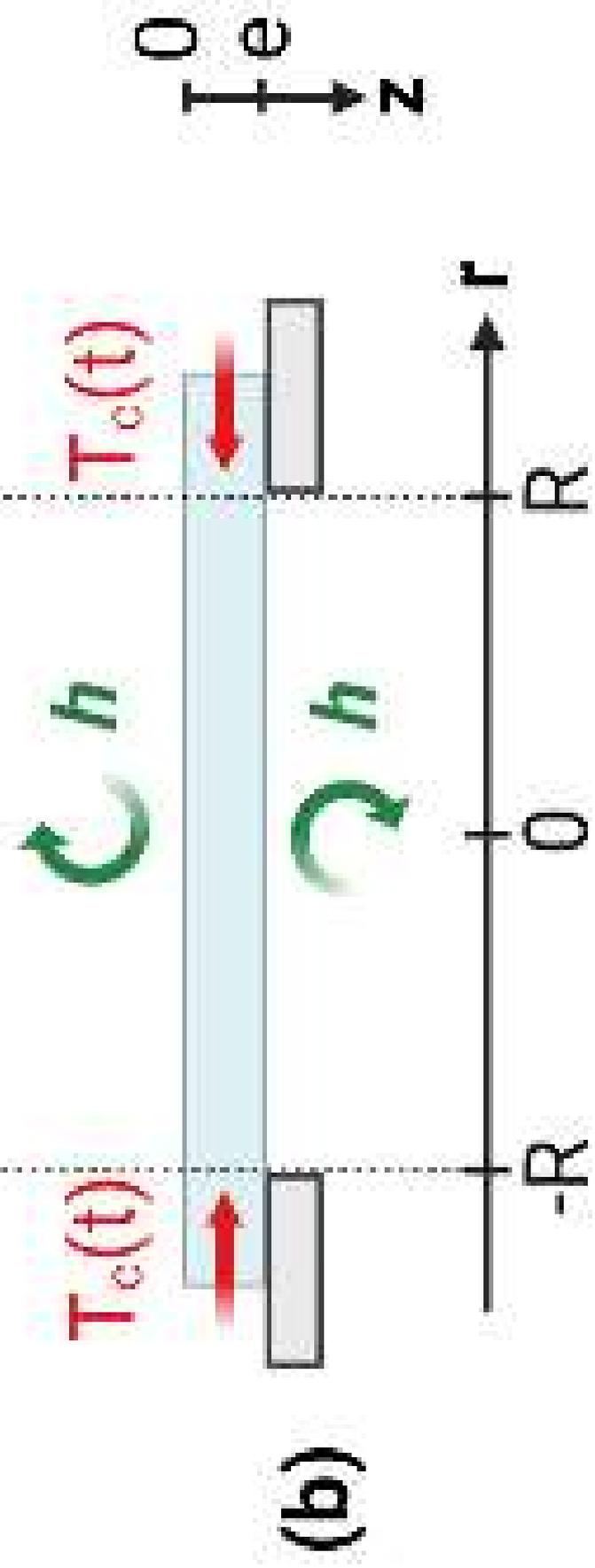

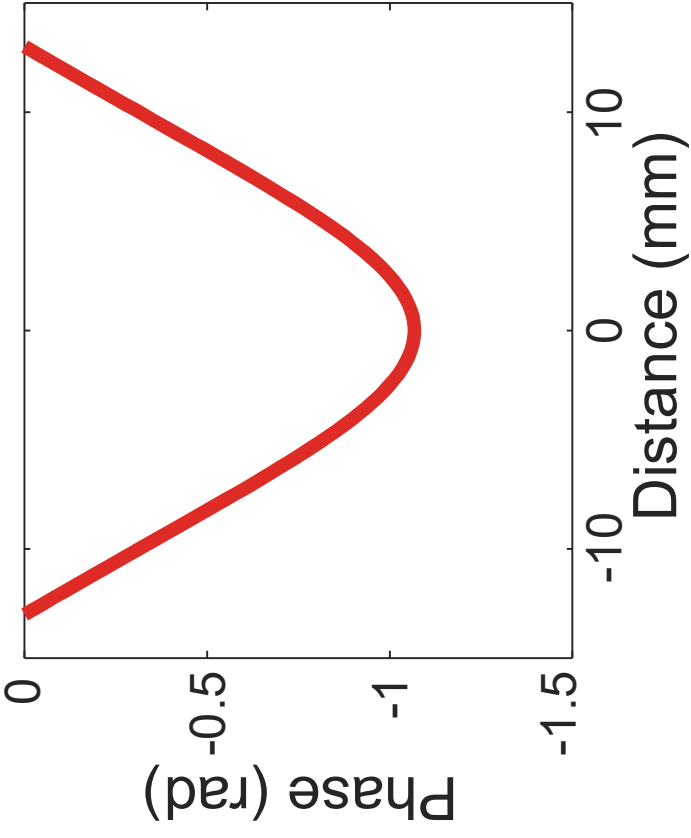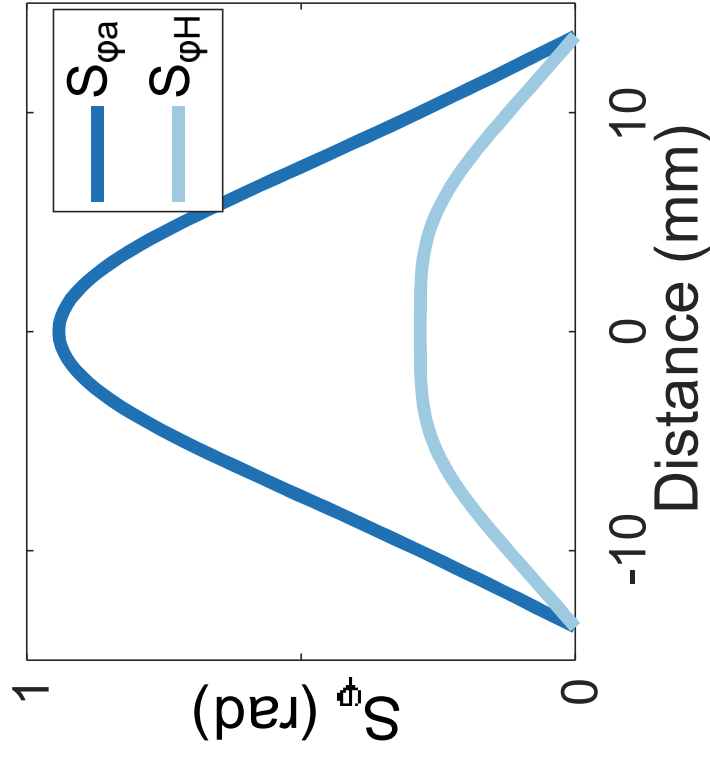
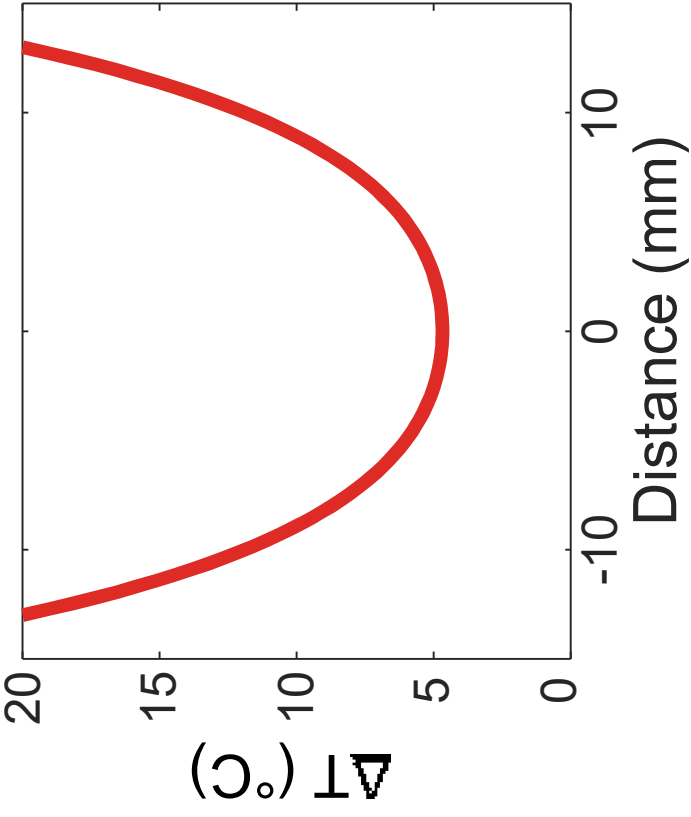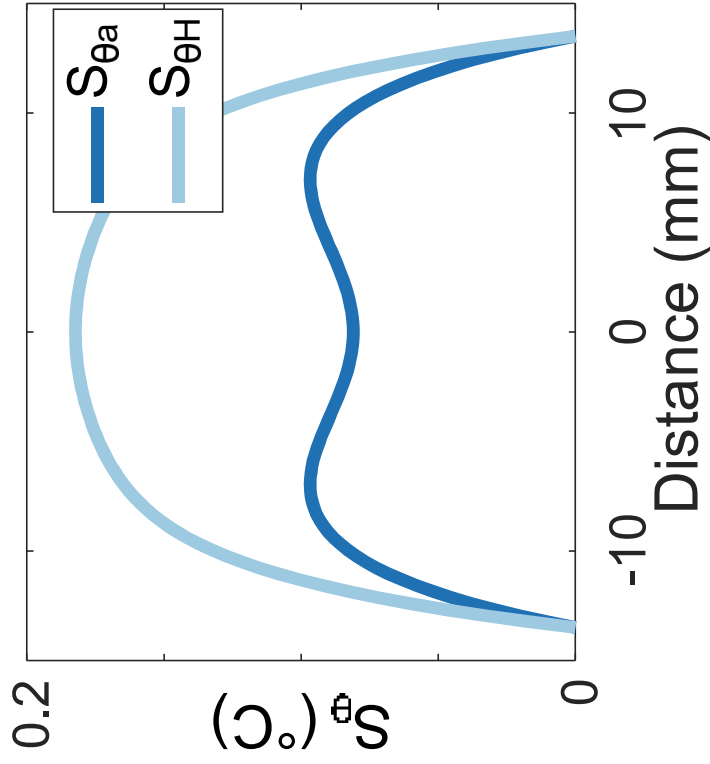

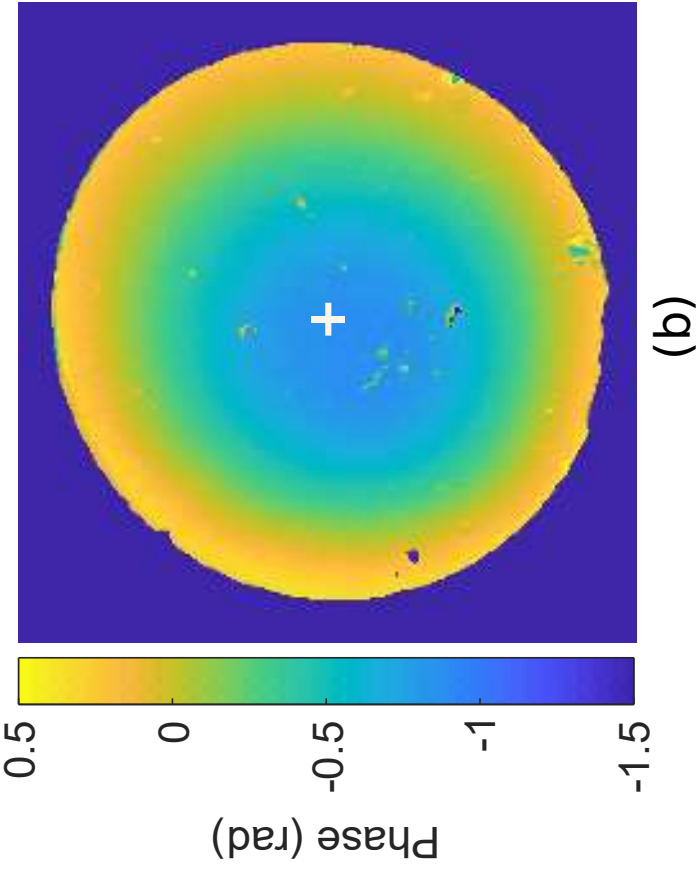
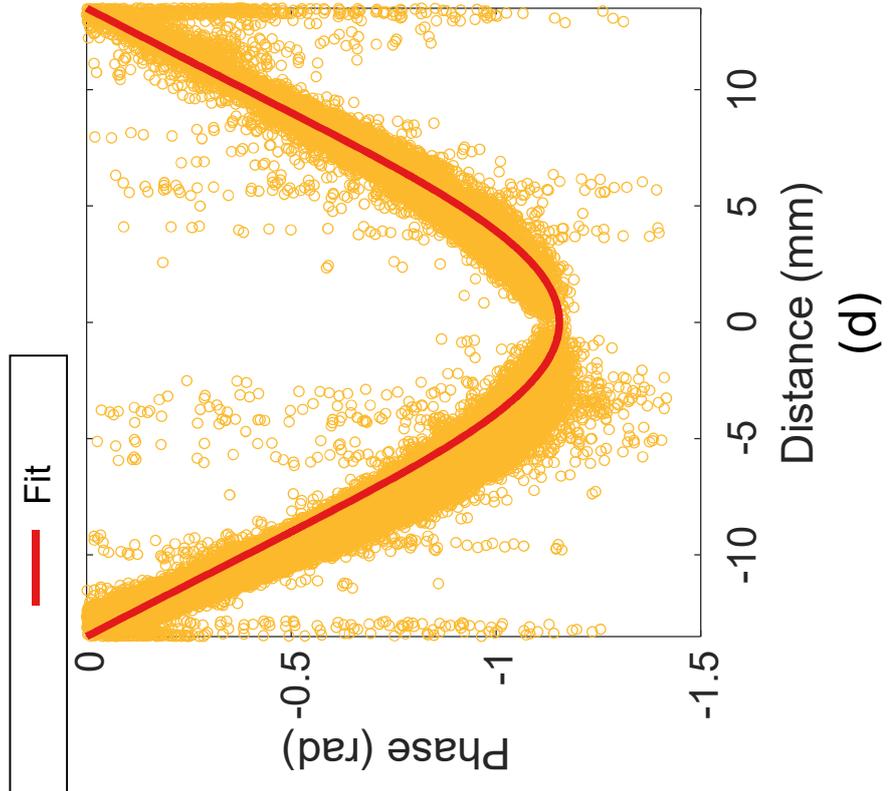
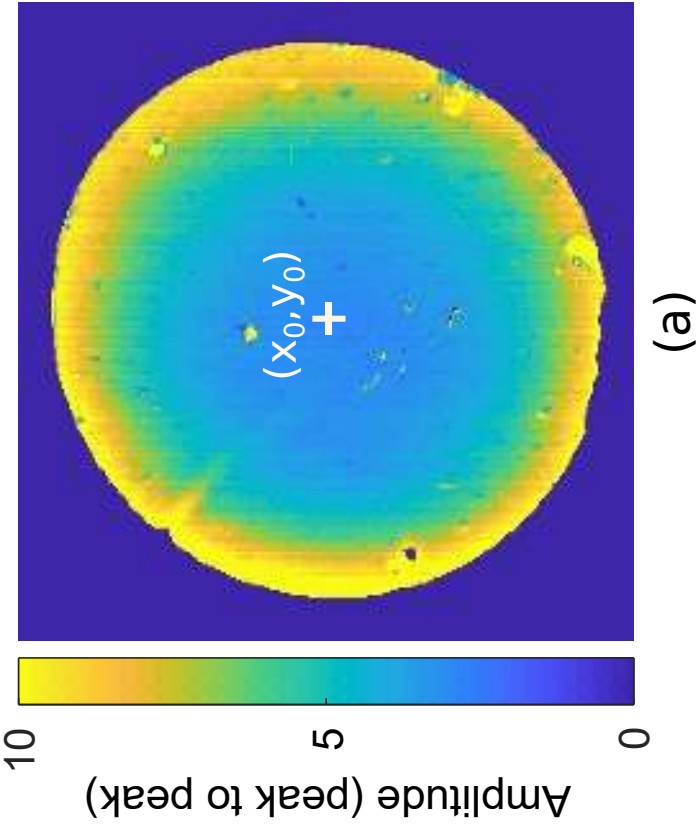
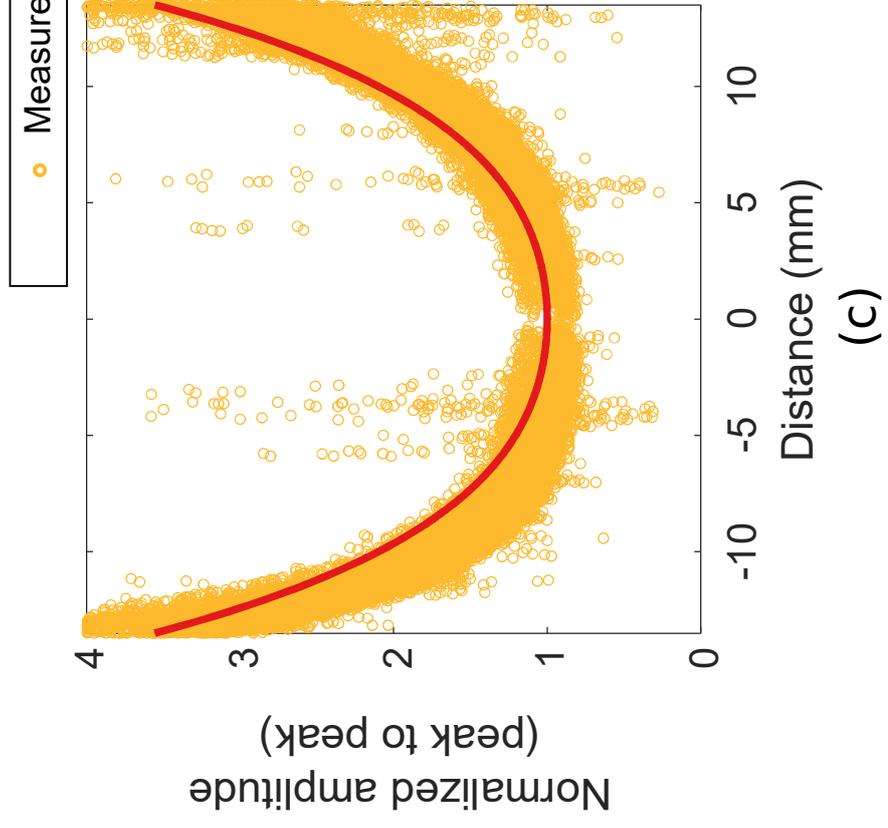

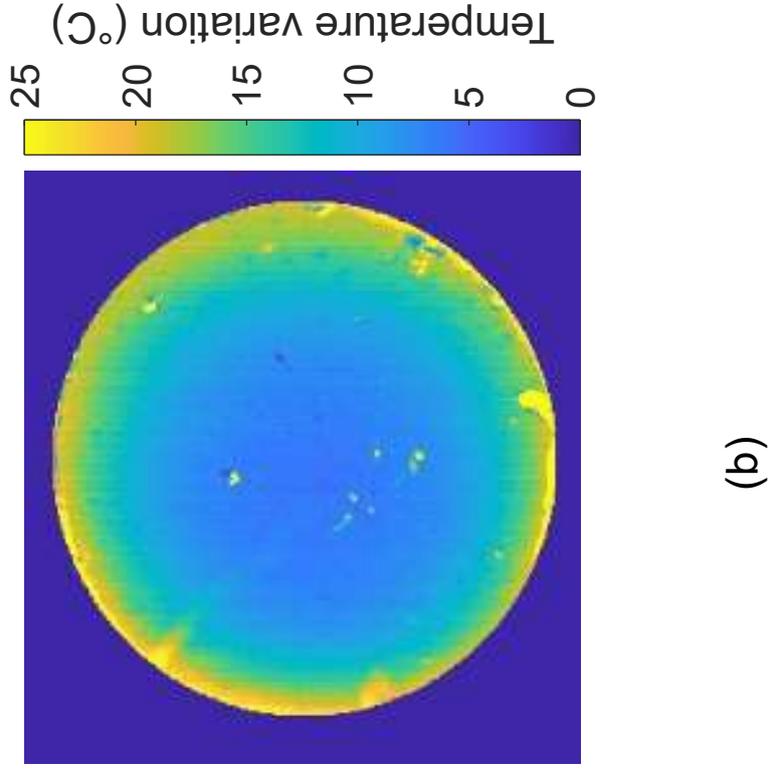
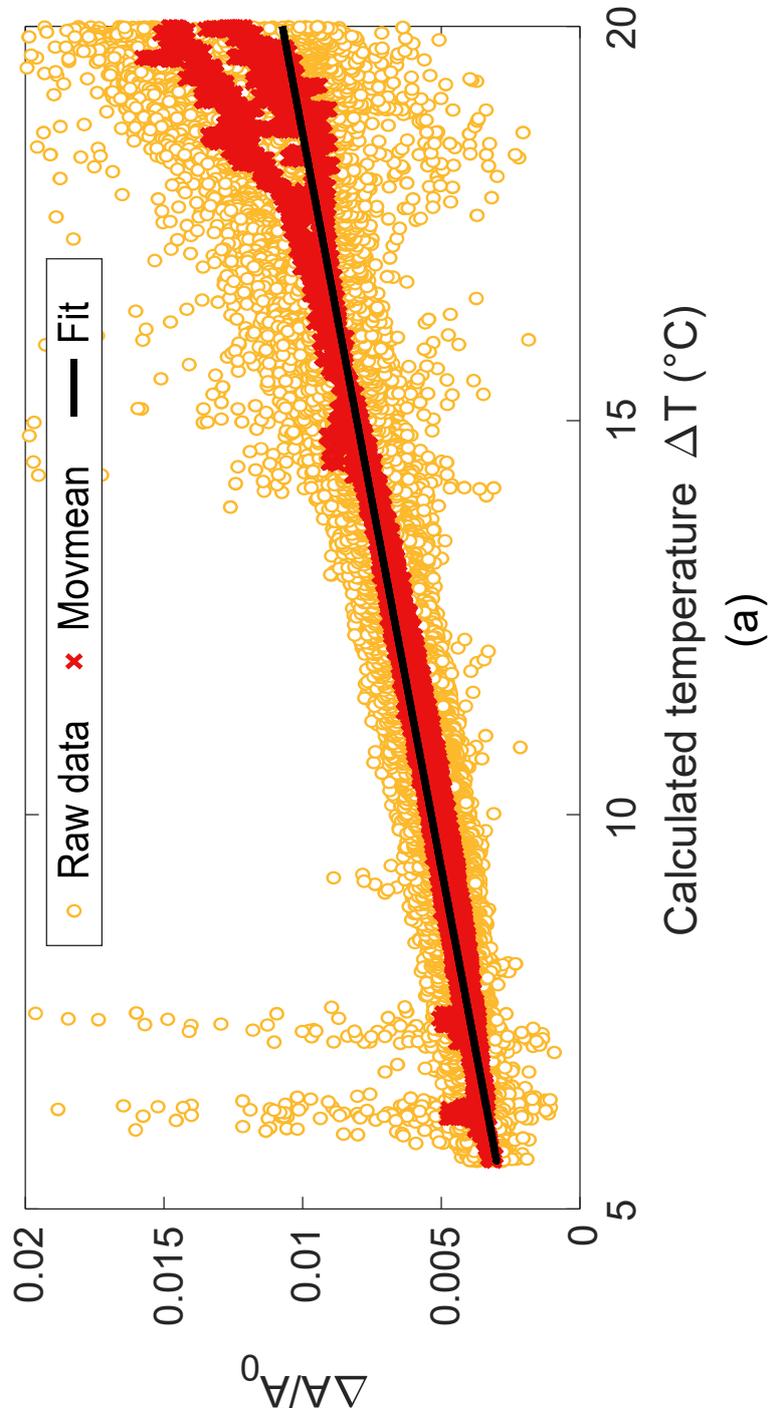